\documentclass{article}

\usepackage[eandd,preprint,nonatbib]{neurips_2026}

\usepackage{fontspec}
\usepackage{graphicx}
\usepackage{booktabs}
\usepackage{multirow}
\usepackage{amsmath}
\usepackage{xurl}
\usepackage[table]{xcolor}
\usepackage{csquotes}
\usepackage{microtype}

\usepackage[style=authoryear-comp,backend=biber,maxcitenames=2,
            maxbibnames=6,minbibnames=1,doi=false,url=false,
            isbn=false,eprint=true]{biblatex}
\AtEveryBibitem{%
  \clearfield{note}%
  \clearfield{annote}%
  \clearfield{annotation}%
}

\newfontfamily\pashtofont[
  Script=Arabic,
  Scale=1.0
]{Amiri-Regular.ttf}[BoldFont=Amiri-Bold.ttf]

\usepackage{newunicodechar}
\newunicodechar{ټ}{{\pashtofont ټ}}
\newunicodechar{ډ}{{\pashtofont ډ}}
\newunicodechar{ڼ}{{\pashtofont ڼ}}
\newunicodechar{ړ}{{\pashtofont ړ}}
\newunicodechar{ښ}{{\pashtofont ښ}}
\newunicodechar{ږ}{{\pashtofont ږ}}
\newunicodechar{ځ}{{\pashtofont ځ}}
\newunicodechar{ۍ}{{\pashtofont ۍ}}

\title{Script collapse in multilingual ASR:\\
A reference-free metric and 100-pair benchmark}

\author{%
  Hanif Rahman\\
  Pashto DAO
}

\begin{document}

\maketitle

\begin{abstract}
% ~170 words — 4 sentences: problem → gap → contribution → finding
% WRITING RULES: no banned words, no -ing tail clauses, sentence case.

Word error rate (WER) is the dominant metric for automatic speech recognition,
yet it cannot detect a systematic failure mode: models that produce fluent output
in the wrong writing system.
We define \emph{Script Fidelity Rate} (SFR), the fraction of hypothesis
characters in the target script block, computable without reference transcriptions,
and report a systematic measurement of script collapse across ten
languages spanning six writing systems and ten models (seven Whisper sizes,
MMS-1B, SeamlessM4T-v2, and Gemma~4~E2B) on FLEURS test sets.
Across 100 evaluated model--language pairs, 21 (21\,\%; 95\,\% Wilson CI:
14--30\,\%) exhibit script collapse (SFR $<$ 10\,\%): 20 involve Whisper and
one involves Gemma~4~E2B on Urdu under a generic transcription prompt.
In a ten-language Gemma~4 probe, script-aware prompting raises mean SFR from
71.2\,\% to 97.7\,\%, fixes Urdu collapse (6.5\,\% to 97.0\,\%), and recovers
5.9 chrF on downstream NLLB translation for the six languages whose baseline
SFR is below 90\,\%.
We identify four collapse patterns: Latin phonetic substitution, Arabic
substitution for Somali, Devanagari substitution for Bengali/Malayalam, and
unique-script Latin collapse for Georgian.

\end{abstract}

% §1 Introduction — ~0.75 pages in IEEE SPL 5-page two-column format
% Goal: establish the gap, motivate SFR, preview contributions.
%
% WRITING RULES: no banned words, no -ing tail clauses, no title-case headings.

\section{Introduction}
\label{sec:intro}

An ASR system can achieve state-of-the-art word error rate while producing output
that no speaker of the target language can read.
Word error rate treats all substitutions equally regardless of which writing system
they appear in.
A Whisper model forced to transcribe Pashto audio can output fluent Arabic or
Latin text and still register a finite WER, because WER only counts word edits;
it does not verify that the output is in the language's standard orthography.
We call this \emph{script collapse}: the model's decoder abandons the target
script entirely, producing text that is phonetically plausible in some
writing system but orthographically unusable.

The benchmark below measures this directly for Pashto and other target
languages: for several Whisper sizes, output remains finite under WER while SFR
shows that the model has moved into Latin or another substitute script.
Informal GitHub discussions document the same phenomenon in Hindi
(\texttt{openai/whisper\#1662}), Somali (\texttt{\#234}), Malayalam
(\texttt{\#1019}), and Bengali (\texttt{\#203}).
We found no peer-reviewed paper that formally defines a script-fidelity metric,
measures it across languages and models, and identifies the conditions under
which WER is misleading without it.

\paragraph{Related work.}
\textcite{Manohar2024Normalization} document normalisation
artefacts that inflate WER for non-Latin scripts.
\textcite{AdvocateCER2025} argue for character error rate as a complement
to WER for Indic languages.
\textcite{Bandarupalli2025PeroArabic} report Whisper WER on Urdu
but do not measure script output rates.
None of these papers defines or measures script fidelity as a first-class metric.

Script collapse is a specific form of decoder hallucination in sequence-to-sequence
models: the decoder generates fluent, well-formed output in the wrong writing system.
The hallucination literature for neural speech models examines phantom insertions,
repetition, and language confusion artefacts arising from training data
distribution~\cite{Radford2023Whisper}, but does not define a scalar metric for
script-level failure.

Two existing research directions are related but distinct.
\emph{Language identification} (LID) from ASR output classifies the output
language using a trained classifier over hypothesis text or audio features,
producing a categorical label.
SFR differs in three respects: (1) it produces a continuous score in $[0,1]$,
not a categorical label; (2) it requires no trained classifier, only Unicode
block membership; and (3) it detects failures that LID cannot, for example
Devanagari output on a Bengali utterance, where a language-level classifier
would report an identical ``Indic language'' label for both the correct and
the collapsed output.
\emph{Script detection} in NLP preprocessing pipelines identifies the script of
a text string for tokenisation or downstream routing; it is not an evaluation
metric and makes no comparison to an expected target.
SFR repurposes the same Unicode block lookup as a scalar evaluation metric
tied to a specific target language.
This conditioning on the expected target script is the difference between
ordinary script detection and SFR: a script detector can label a Bengali
hypothesis as Devanagari, but SFR turns that label into an ASR evaluation signal
by asking whether Devanagari is the expected output script for Bengali.
Corpus-level aggregation then makes the failure measurable across a benchmark
and comparable to WER.

Character-level Unicode block membership is a standard operation in multilingual
text processing~\cite{Manohar2024Normalization, Bandarupalli2025PeroArabic,
Sproat2016text}: the script of any character is determined in $O(1)$ from its
code point.
Prior work applies this to text normalisation and data-pipeline preprocessing,
not to ASR evaluation.
Our literature search found no prior treatment of Unicode block analysis as an
ASR evaluation metric in Interspeech, ICASSP, ACL, EMNLP, or IEEE SPL/TASLP
proceedings through 2025.
SFR does not compete with WER or CER; it identifies a failure mode neither
metric can detect.
Although our evaluation demonstrates this failure mode primarily in Whisper, SFR
is architecture-agnostic: it applies equally to any sequence-to-sequence or
CTC-based ASR system whose decoder could in principle produce characters outside
the target script.
Like WER and chrF, SFR is simple by design: it isolates a failure mode that
existing scores fold into general transcription error, then lets us measure and
mitigate that failure across model families.

\paragraph{Contributions.}
This paper makes three contributions:
\begin{enumerate}
  \item A formal definition of Script Fidelity Rate (SFR) as a reference-free
    ASR evaluation metric: it requires only the hypothesis string and a target
    language identifier, making it computable in production deployments without
    labelled data (\S\ref{sec:metric}).
  \item A systematic measurement of SFR across ten models (seven Whisper
    sizes, MMS-1B, SeamlessM4T-v2, and Gemma~4~E2B) and ten languages on FLEURS
    test sets, exposing where and how script collapse occurs (\S\ref{sec:results}).
  \item A mitigation and downstream validation for an instruction-following
    speech model: script-aware prompting raises Gemma~4 mean SFR from
    71.2\,\% to 97.7\,\%, and downstream NLLB translation recovers most on the
    languages where SFR improves most (\S\ref{sec:results}).
\end{enumerate}

Across 100 evaluated model--language pairs, 21 (21\,\%) exhibit script collapse
(SFR $<$ 10\,\%).
Twenty involve Whisper; one is Gemma~4~E2B on Urdu under a generic transcription
prompt.
We identify four distinct collapse patterns: Latin phonetic output (smaller
Whisper on Indic), Arabic for Somali's Latin orthography, Devanagari for
Bengali/Malayalam (larger Whisper treats all Indic audio as Hindi), and
Latin output for Georgian, a script with no phonetic relationship to Latin.

% §2 Script Fidelity Rate metric — ~0.7 pages
% Goal: formal definition, Unicode block table, validation protocol.

\section{Script Fidelity Rate}
\label{sec:metric}

\subsection{Definition}

Let $H = c_1 c_2 \dots c_n$ be an ASR hypothesis string after Unicode NFC
normalisation.
Define the \emph{countable characters} of $H$ as those that are neither
whitespace, punctuation (Unicode general category P*), nor formatting
characters (category C*):
\[
  \hat{H} = \{ c_i \in H \mid c_i \notin \text{whitespace} \cup \text{punct} \cup \text{format} \}
\]

For a target language $\ell$ with a designated script $\mathcal{S}_\ell$
defined by a set of Unicode code-point ranges $R_\ell$ and an optional set of
language-unique code points $U_\ell$, let
$n_\ell(H) = |\{ c \in \hat{H} \mid \text{ord}(c) \in R_\ell \text{ or } c \in U_\ell \}|$
be the count of target-script characters.
The \emph{Script Fidelity Rate} of $H$ is:
\[
  \text{SFR}(H, \ell) =
  \begin{cases}
    n_\ell(H) \;/\; |\hat{H}| & |\hat{H}| > 0 \\[4pt]
    \texttt{null}              & |\hat{H}| = 0
  \end{cases}
\]

Corpus-level SFR is the mean over non-null utterance-level values.
A value of 1.0 means every output character is in the target script; a value
near 0 means the model produced output entirely in a different writing system,
a condition we term \emph{script collapse}.

\subsection{Metric properties}

SFR satisfies two basic desiderata for an evaluation metric:

\begin{enumerate}
  \item \textbf{Boundedness.} $\text{SFR}(H, \ell) \in [0, 1]$ for any non-null
    hypothesis, since $n_\ell(H) \leq |\hat{H}|$ by definition.
  \item \textbf{Monotonicity.} Replacing any non-target-script character in $H$
    with a target-script character cannot decrease SFR, and strictly increases it
    when $|\hat{H}| > 0$.
\end{enumerate}

We report the unweighted mean over utterances so each test item contributes
equally to the corpus score.
A character-weighted variant is possible but is not used in this benchmark.

SFR is not a replacement for WER: it is a precondition check.
A WER value is interpretable only after SFR confirms the output is in the target
script.

\subsection{Reference-free property}

SFR requires only the hypothesis $H$ and the target language identifier $\ell$.
No reference transcription is needed.
This distinguishes SFR from every standard ASR metric (WER, CER, MER): those
metrics require labelled ground truth, which is unavailable in production
deployments.
SFR can therefore be computed as a continuous deployment audit. It flags script
collapse before users report unintelligible output.

\subsection{Unicode block specifications}

Table~\ref{tab:scripts} lists the Unicode block ranges and unique code points
used for each language in this study.
For Pashto, the unique-code-point set $U_\ell$ (glyphs absent from standard
Arabic and Urdu) provides an unambiguous positive signal even when the Arabic
block is shared with Urdu, Dari, and other Perso-Arabic languages.
For Somali, the target script is Latin; the main failure mode is Arabic output
on Somali audio.

\begin{table}[htb]
\centering
\caption{Script configurations for all ten target languages.
  $R_\ell$ = primary Unicode block range(s); $U_\ell$ = unique code points
  (non-empty for languages sharing a block with others).}
\label{tab:scripts}
\begin{tabular}{llll}
\toprule
Language & Script & $R_\ell$ (main range) & $|U_\ell|$ \\
\midrule
Pashto    & Perso-Arabic & U+0600--06FF          & 12 \\
Urdu      & Perso-Arabic & U+0600--06FF          &  5 \\
Arabic    & Perso-Arabic & U+0600--06FF          &  0 \\
Persian   & Perso-Arabic & U+0600--06FF          &  4 \\
Hindi     & Devanagari   & U+0900--097F          &  0 \\
Bengali   & Bengali      & U+0980--09FF          &  0 \\
Malayalam & Malayalam    & U+0D00--0D7F          &  0 \\
Tamil     & Tamil        & U+0B80--0BFF          &  0 \\
Somali    & Latin        & U+0041--005A, 0061--007A &  0 \\
Georgian  & Georgian     & U+10A0--10FF          &  0 \\
\bottomrule
\end{tabular}
\end{table}

\subsection{Relationship to WER}

SFR is a script-support statistic, not a distance metric.
WER and CER compare a hypothesis with a reference after normalisation; SFR asks
whether the hypothesis lies in the target writing system before that comparison.
This distinction makes SFR reference-free: it measures a property of the
hypothesis and the target language specification, not edit distance to a labelled
transcription.
For shared-script families, SFR is an upper bound on language-specific
orthographic fidelity: it can verify that output uses the Perso-Arabic block, but
not by itself that the words are Pashto rather than Urdu or Arabic.

Proposition 1 (independence from WER).
There exist hypotheses $H,H'$ and a reference $R$ such that
$\mathrm{WER}(H,R)=\mathrm{WER}(H',R)$ but
$\mathrm{SFR}(H,\ell)\neq\mathrm{SFR}(H',\ell)$.
Conversely, there exist hypotheses with equal SFR and unequal WER.

\emph{Proof.}
Let $\ell$ be Pashto and let $R=(r_1,\ldots,r_m)$ be a Pashto-script reference
with $m\geq2$ tokens.
Let $H_L=(x,r_2,\ldots,r_m)$ where $x$ is a Latin token, and let
$H_P=(y,r_2,\ldots,r_m)$ where $y$ is a Pashto-script token different from
$r_1$.
Both hypotheses have one substitution, so both have WER\,=\,$1/m$.
However, $\mathrm{SFR}(H_L,\ell)<1$ because $x$ is Latin, while
$\mathrm{SFR}(H_P,\ell)=1$.
Conversely, $R$ and any all-Pashto-script hypothesis with no token overlap with
$R$ have equal SFR but different WER.

Thus low SFR does not make WER undefined; it means WER lacks the information
needed to distinguish wrong-script output from same-script lexical error.
SFR is therefore a validity check for interpreting WER, not a replacement for WER.
In the Bengali example discussed in \S\ref{sec:discussion}, Whisper large-v2 has
WER\,=\,113\,\% and SFR\,=\,0.7\,\%: WER reports a bad transcript, while SFR
shows that the transcript is in the wrong script.

\subsection{Failure taxonomy}

We distinguish two script-fidelity failure modes, identified empirically in
\S\ref{sec:results}:

\begin{enumerate}
  \item \textbf{Script substitution.} The model outputs valid text in a
    different writing system (e.g.\ Latin transliteration, Arabic for a
    Devanagari language, or English text). This is the dominant failure mode
    for Whisper on non-Latin-script languages and the defining characteristic
    of script collapse.
  \item \textbf{Diacritic stripping.} The model outputs characters in the
    correct Unicode block but omits diacritics obligatory in that orthography.
    SFR remains high but lexical accuracy degrades. This is the predominant
    failure mode for MMS on Indic languages.
\end{enumerate}

A third failure mode, \emph{decoder looping} (repetition of a short phrase,
producing very high WER while SFR may remain high), is visible in the data
but is not a script-fidelity problem: the output is in the correct script.
We note it here for completeness and discuss the clearest instance (Somali,
Whisper tiny) in \S\ref{sec:results}.

\subsection{Validation protocol}

Before running any model, the SFR implementation is validated against known
positives and negatives for each language (see \texttt{scripts/script\_fidelity.py}).
For Pashto, the validation includes examples with Pashto-unique glyphs and Latin
negative controls.

% §3 Experimental setup — ~0.5 pages

\section{Experimental setup}
\label{sec:setup}

\subsection{Datasets}

We evaluate on FLEURS~\cite{Conneau2022FLEURS} test splits for ten languages
covering six writing systems and three roles.
\emph{Collapse targets} are languages where script collapse is expected based on
training data sparsity: Pashto, Hindi, Bengali, Malayalam, Somali, and Georgian.
\emph{Script-family controls} are Perso-Arabic languages with abundant training
data (Arabic, Persian, Urdu) where SFR should remain near 100\,\% regardless of
model size, validating that the metric does not penalise correct output.
\emph{Dravidian extension}: Tamil adds a second Dravidian script family to
Bengali and Malayalam, testing whether collapse is script-specific or
architecture-specific.
All ten FLEURS test splits contain at least 250 utterances ($\min = 299$ for Urdu).
Table~\ref{tab:datasets} lists the evaluation sets with utterance counts.

\begin{table}[htb]
\centering
\caption{Evaluation datasets. $N$ = FLEURS test utterances. Role: C = collapse target, V = validation control, E = extension.}
\label{tab:datasets}
\begin{tabular}{lllrr}
\toprule
Language  & FLEURS code & Script       & $N$ & Role \\
\midrule
Pashto    & \texttt{ps\_af} & Perso-Arabic &  512 & C \\
Urdu      & \texttt{ur\_pk} & Perso-Arabic &  299 & V \\
Arabic    & \texttt{ar\_eg} & Perso-Arabic &  428 & V \\
Persian   & \texttt{fa\_ir} & Perso-Arabic &  871 & V \\
Hindi     & \texttt{hi\_in} & Devanagari   &  418 & C \\
Bengali   & \texttt{bn\_in} & Bengali      &  920 & C \\
Malayalam & \texttt{ml\_in} & Malayalam    &  958 & C \\
Tamil     & \texttt{ta\_in} & Tamil        &  591 & E \\
Somali    & \texttt{so\_so} & Latin        & 1019 & C \\
Georgian  & \texttt{ka\_ge} & Georgian     &  979 & C \\
\bottomrule
\end{tabular}
\end{table}

\subsection{Models}

We evaluate ten models spanning three architectures and four training paradigms:
seven Whisper sizes, MMS-1B, SeamlessM4T-v2, and one multimodal LLM.

\paragraph{Whisper~\cite{Radford2023Whisper}.}
Seven sizes: tiny, base, small, medium, large-v2, large-v3, and large-v3-turbo.
Inference uses the HuggingFace \texttt{transformers} pipeline with the language
token forced to the target language and greedy decoding (\texttt{num\_beams=1}).
Greedy decoding is used consistently across all models to remove beam-search
hyperparameters as a confound.
We do not test alternative decoding strategies in this paper.

\paragraph{MMS-1B~\cite{Pratap2023MMS}.}
Meta's massively multilingual CTC model trained on over 1,100 languages via
language-specific adapters.
MMS-1B was evaluated on all ten languages.

\paragraph{SeamlessM4T-v2-large~\cite{Barrault2023SeamlessM4T}.}
Meta's multilingual speech-to-text model evaluated with forced target language
using FLORES-200 language codes.

\paragraph{Gemma~4~E2B~\cite{Riviere2025Gemma4}.}
Google DeepMind's instruction-tuned multimodal LLM
(E2B: 2.3\,B effective / 5.1\,B total parameters;
\texttt{unsloth/gemma-4-E2B-it}, Apache-2.0).
Transcription is produced via instruction following rather than forced decoding:
the audio is tokenised and the model generates a hypothesis given a text prompt.
We evaluate Gemma~4 on all ten languages.
The baseline prompt asks the model to transcribe the speech segment in its
original language, with digit-formatting instructions.

\paragraph{Script-aware prompting and downstream MT.}
For Gemma~4 only, we run a paired mitigation experiment with the same decoding
settings and a script-aware prompt that names the language and target script.
We then translate gold FLEURS transcripts, Gemma baseline transcripts, and
script-aware transcripts to English using NLLB-200 distilled 600M~\cite{NLLB2022}.
The English FLEURS test transcript is the reference.
We score corpus chrF and BLEU for each language and compare the ASR-fed MT
outputs with the gold-transcript MT control.

\subsection{Text normalisation}

WER counts insertions, deletions, and substitutions relative to reference length,
so excessive insertions, produced for example by decoder looping, can push
WER above 100\,\%.

WER and CER are computed after language-specific normalisation:
Arabic-script languages (Pashto, Urdu, Arabic, Persian) strip diacritics,
tatweel, and punctuation; Indic and Tamil strip punctuation and Indic digits;
Somali lowercases and strips punctuation; Georgian lowercases (converting rare
Asomtavruli capitals to standard Mkhedruli) and strips punctuation.
SFR is computed on the \emph{raw, unnormalized} hypothesis: normalisation can
alter Unicode code points and artificially inflate SFR.

\subsection{Compute}

Whisper and MMS run on a single NVIDIA A40 (48\,GB VRAM, RunPod).
SeamlessM4T-v2-large runs on a single NVIDIA RTX\,4090 (24\,GB VRAM).
Gemma~4~E2B runs on an Apple M4 Max (40\,GB unified memory) using the
\texttt{bfloat16} MPS backend.
Gemma~4 script-aware prompting and NLLB downstream MT use the same machine with
deterministic decoding.
Whisper, MMS, and SeamlessM4T use \texttt{float16} precision on CUDA.
Results and per-utterance prediction files are included in the anonymous
supplementary material.

% §4 Results — ~1.2 pages; main empirical contribution

\section{Results}
\label{sec:results}

\subsection{Script Fidelity Rate matrix}

\begin{figure}[htb]
  \centering
  \includegraphics[width=\linewidth]{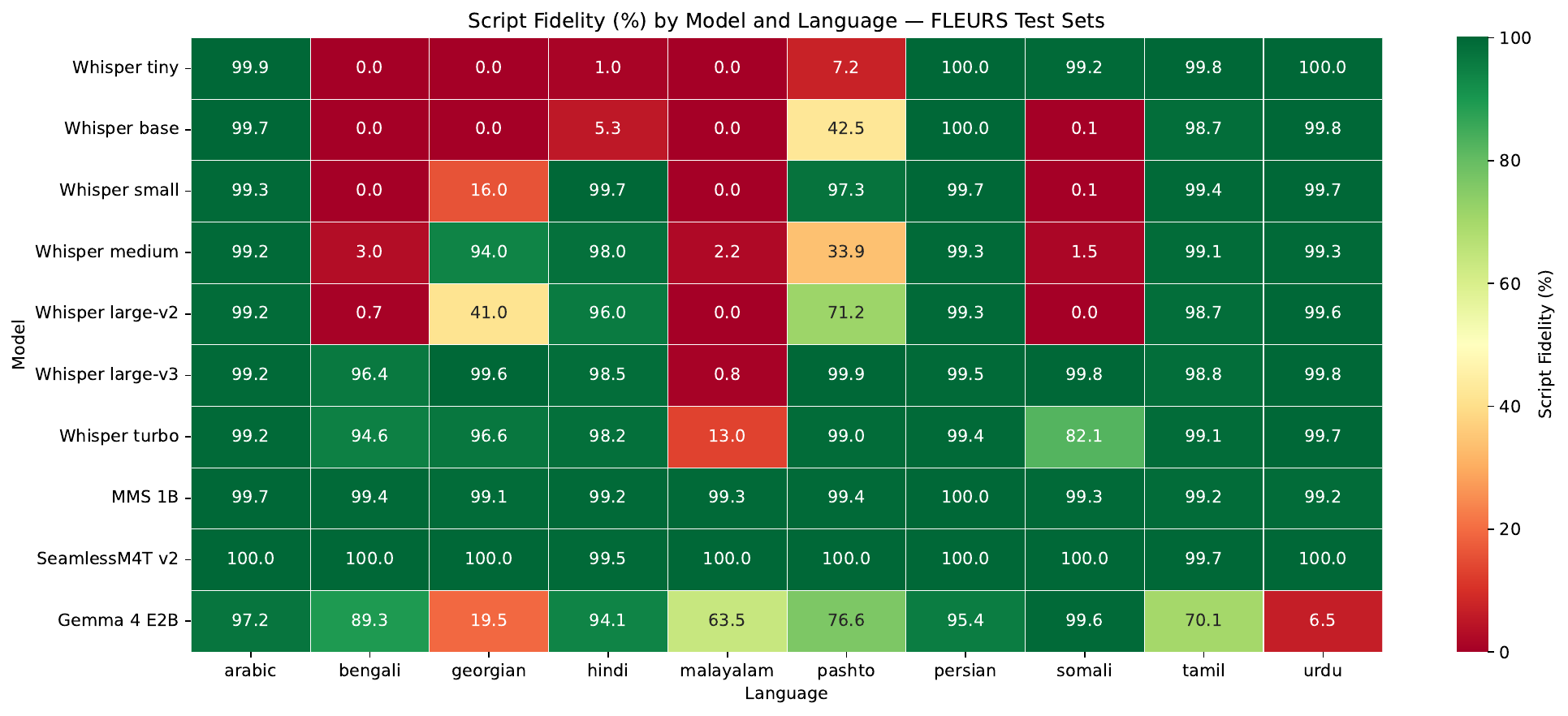}
  \caption{Script Fidelity Rate (\%) for the 100 evaluated model--language pairs
    on FLEURS test sets.
    Green cells indicate high script fidelity; red cells indicate script
    collapse.
    The full numeric matrix appears in Appendix~\ref{app:matrix}.}
  \label{fig:sfr_heatmap}
\end{figure}

Figure~\ref{fig:sfr_heatmap} reports SFR for all model--language pairs.
Appendix~\ref{app:matrix} gives the same values as a numeric table.
The WER-vs-SFR scatter plot for all pairs appears in Appendix~\ref{app:scatter}.

Twenty-one of 100 evaluated pairs (21\,\%; 95\,\% Wilson CI: 14--30\,\%) exhibit
script collapse.
Twenty involve Whisper; the remaining collapse case is Gemma~4~E2B on Urdu
under a generic instruction-following transcription prompt.
MMS-1B and SeamlessM4T-v2 show no collapse cases.
The three Perso-Arabic validation languages (Arabic, Persian, Urdu) show no
collapse for Whisper, MMS-1B, and SeamlessM4T-v2, confirming that SFR does not
penalise correct output in well-resourced script families.
Tamil shows no collapse either, extending this result to a second Dravidian family.

The 10\,\% collapse threshold is grounded in the observed SFR distribution,
which is strongly bimodal: 21 values fall below 10\,\%, 67 fall above 90\,\%,
and only 12 lie in the intermediate range (13--89\,\%).
The highest collapsed value is 7.2\,\% (Whisper tiny on Pashto) and the lowest
non-collapsed value is 13.0\,\% (Whisper turbo on Malayalam), leaving a natural
gap of 5.8 percentage points.
Any threshold in the interval $[7.2\,\%, 13.0\,\%]$ yields the same 21
collapse cases; the set of collapsed pairs is insensitive to the specific
threshold chosen within this range.
The bimodal structure holds across all four model families: 88 of 100 pairs fall
above 90\,\% or below 10\,\%, with only 12 intermediate values.
Table~\ref{tab:sfr_summary} summarises the SFR distribution by model family.
Appendix~\ref{app:threshold_sensitivity} shows that the main conclusion is
stable under nearby thresholds.

\begin{table}[htb]
\centering
\caption{SFR distribution summary across 100 evaluated model--language pairs.
  Brackets give 95\,\% bootstrap confidence intervals for mean SFR, resampling
  model--language pairs within each family.}
\label{tab:sfr_summary}
\begin{tabular}{lrrr}
\toprule
Model family & Mean SFR & Median & Collapsed \\
\midrule
Whisper       & 65.3 [54.6, 75.5] & 98.3 & 20 / 70 \\
MMS-1B        & 99.4 [99.2, 99.5] & 99.3 &  0 / 10 \\
SeamlessM4T   & 99.9 [99.8, 100.0] & 100.0 & 0 / 10 \\
Gemma 4       & 71.2 [50.2, 88.7] & 82.9 & 1 / 10  \\
\midrule
All models    & 72.8 [64.5, 80.5] & 99.1 & 21 / 100 \\
\bottomrule
\end{tabular}
\end{table}

Script collapse appears across all seven Whisper sizes.
Even Whisper large-v3 collapses on Malayalam (SFR\,=\,0.8\,\%).
Urdu, Arabic, Persian, and Tamil do not collapse for Whisper, suggesting strong
script priors for these higher-resource or more stable orthographies.

Georgian shows a non-monotonic pattern across Whisper sizes.
Whisper tiny and base produce SFR\,=\,0\,\%; 966 of 979 base-model utterances
are Latin-dominant.
Whisper small recovers partially (SFR\,=\,16\,\%), and medium reaches 94\,\%.
Whisper large-v2 then \emph{regresses} to SFR\,=\,41\,\%, before large-v3 recovers
to 99.6\,\%.
This reversal between large-v2 and large-v3 is the sharpest version-to-version
SFR shift in the benchmark and is not reflected in WER: both versions report
WER above 68\,\% on Georgian, giving no signal of the script difference.
The pattern suggests a model-version change in Georgian script coverage, but the
training data change is not publicly documented.

One result sits outside the script-collapse regime: Whisper tiny on Somali
achieves SFR\,=\,99.2\,\% (correct Latin script) but WER\,=\,458\,\%.
This is the decoder-looping failure mode noted in \S\ref{sec:metric}: the model
repeats a short phrase or single token, massively inflating the insertion count
without changing the output script.
SFR correctly reports no script collapse here; the pathological WER is a separate
quality failure that WER itself captures.

\subsection{The script collapse quadrant}

The WER-vs-SFR scatter (Appendix~\ref{app:scatter}) plots all model--language pairs.
The scatter separates script fidelity from lexical accuracy.
The most important region is high WER / low SFR: the output is orthographically
unusable, but WER alone only reports poor accuracy.
For example, Whisper large-v2 on Bengali has WER\,=\,113\,\% while outputting
Devanagari; SFR\,=\,0.7\,\% reveals the wrong-script failure.
High WER / high SFR is different: MMS-1B and SeamlessM4T occupy this region on
harder languages, where the script is correct but lexical accuracy remains low.
No evaluated pair falls in the low-WER / low-SFR quadrant.

% Scatter plot moved to appendix to stay within the 9-page limit.
% The appendix is not counted toward the page limit per NeurIPS E&D rules.

\subsection{Script-aware prompting and downstream MT}
\label{sec:mitigation}

For Gemma~4, script-aware prompting is a lightweight mitigation.
It raises mean SFR from 71.2\,\% to 97.7\,\% across the ten languages and
recovers all six baseline languages with SFR below 90\,\% into the high-SFR
region.
The largest single change is Urdu: the generic prompt produces Devanagari
dominant output (SFR\,=\,6.5\,\%), while the script-aware prompt recovers Urdu
Perso-Arabic output (SFR\,=\,97.0\,\%).

We also test whether SFR changes matter for a downstream text pipeline.
Using English FLEURS transcripts as references, we translate gold transcripts,
Gemma baseline outputs, and script-aware outputs into English with NLLB-200.
The six languages with baseline SFR below 90\,\% recover 5.9 chrF and 3.1 BLEU
on average, while the four high-SFR control languages recover only 0.6 chrF and
0.3 BLEU.
Across languages, the correlation between SFR gain and chrF recovery is strong
(Spearman $\rho=0.93$).

\begin{table}[htb]
\centering\small
\caption{Gemma~4 script-aware prompting and downstream MT recovery.
  $\Delta$WER is script-hint minus baseline, so negative is better.
  $\Delta$chrF and $\Delta$BLEU are script-hint MT minus baseline MT against
  English FLEURS references.}
\label{tab:gemma_mitigation_mt}
\setlength{\tabcolsep}{4pt}
\begin{tabular}{lrrrrr}
\toprule
Language & Base SFR & Hint SFR & $\Delta$WER & $\Delta$chrF & $\Delta$BLEU \\
\midrule
Pashto    & 76.6 & 98.8 & -12.6 &  4.4 & 1.1 \\
Urdu      &  6.5 & 97.0 & -75.3 &  5.0 & 3.9 \\
Bengali   & 89.3 & 97.1 &  -5.0 &  0.7 & 0.8 \\
Malayalam & 63.5 & 98.0 & -23.9 &  6.7 & 4.6 \\
Tamil     & 70.1 & 96.4 & -15.8 &  4.9 & 2.2 \\
Georgian  & 19.5 & 93.0 & -43.4 & 13.5 & 6.2 \\
\midrule
Low-SFR mean & 54.2 & 96.7 & -29.3 & 5.9 & 3.1 \\
High-SFR control mean & 96.6 & 99.2 & -1.8 & 0.6 & 0.3 \\
\bottomrule
\end{tabular}
\end{table}

Prompting is not a complete solution.
It improves WER for nine of ten languages, so the MT gains combine better script
choice with better lexical accuracy.
SFR is still needed because the Somali control stays high-SFR but high-WER, and
some utterances remain mixed-script even after prompting.

\subsection{Failure taxonomy}

Per-utterance dominant-script analysis identifies four collapse patterns with
distinct substitute scripts.
Smaller Whisper models often romanize Indic and Pashto audio; several Whisper
sizes output Arabic script for Somali; larger Whisper models often output
Devanagari for Bengali or Malayalam; and Whisper tiny/base output Latin for
Georgian, a script with no Latin relationship.
Appendix~\ref{app:taxonomy_table} gives the dominant-script breakdown, and
Appendix~\ref{app:taxonomy_details} gives examples and a longer discussion.

% §5 Discussion

\section{Discussion}
\label{sec:discussion}

\subsection{Why WER masks script collapse}

WER is computed over word sequences after normalisation.
When a model outputs Latin transliterations of Hindi, or Devanagari for Bengali,
the normalised reference and hypothesis both consist of space-separated tokens.
In every script-collapse case observed in this study (Table~\ref{tab:results}),
WER ranges from 96\,\% to 396\,\%.
These values indicate poor accuracy, but not which script was produced.
A system reporting WER\,=\,113\,\% on a Bengali test set is indistinguishable
from a partially functional Bengali model at the same accuracy; only
SFR\,=\,0.7\,\% reveals that the output is Devanagari throughout.

For a Unicode text pipeline, script collapse is a silent total failure: the
downstream component receives a sequence with no overlap with the expected
vocabulary for that language.
The bimodal SFR distribution suggests that script failure is usually a decoder
script choice rather than gradual character noise.
Most model--language pairs either commit to the target script or to a substitute
script; intermediate SFR values arise when outputs mix scripts or when only part
of the corpus collapses.

\subsection{Architecture comparison}

Whisper is trained on web data with heavy Latin-script coverage~\cite{Radford2023Whisper}.
Its decoder can acquire a Latin-output prior for non-Latin languages with sparse
training data.
MMS-1B adapters~\cite{Pratap2023MMS} and SeamlessM4T-v2's mBART
decoder~\cite{Barrault2023SeamlessM4T} keep output in the target script
(SFR\,$\geq$\,99\,\% across all languages).

Gemma~4~E2B fails through instruction following rather than forced decoding.
The generic prompt collapses on Urdu and mixes scripts on five other languages;
naming the target script recovers all six Gemma baseline languages with SFR
below 90\,\% to high SFR.
The saved Pashto outputs show the distinction: Gemma baseline outputs drift
across scripts, while the script-aware prompt usually returns Pashto-script text
with occasional Hindi, English, or romanized fragments.
Table~\ref{tab:gemma_mitigation_mt} shows the largest chrF gains on Georgian,
Malayalam, Urdu, Tamil, and Pashto, though WER remains high for several languages.

\subsection{Model-version auditing}

The Georgian results show model-version changes that WER hides.
Whisper large-v2 achieves SFR\,=\,41\,\% on Georgian; large-v3 achieves
SFR\,=\,99.6\,\%.
This may reflect changed training data, tokenization, or decoding priors, but
the public model documentation does not identify the cause.

\subsection{Reference-free audit and LID}

SFR needs only the hypothesis string and target language identifier, so it can be
computed where reference transcriptions do not exist.
LID is a useful second-stage diagnostic, but not a replacement.
On Gemma~4 Urdu, baseline outputs have SFR\,=\,6.5\,\% and are labelled Hindi
by langdetect for 91\,\% of utterances; the script-aware prompt raises SFR to
97.0\,\% and Urdu LID to 98.3\,\%.
Yet langdetect labels high-SFR Georgian script-hint output as Estonian for
83.0\,\% of utterances and labels high-SFR Pashto output mostly Persian because
Pashto is unsupported.
The practical audit is therefore SFR first, then WER/CER, LID, or human review
for substitute-language diagnosis.
Appendix~\ref{app:lid} gives the full SFR+LID summary.

\subsection{Limitations and societal impact}

SFR is a precondition check, not a substitute for WER, CER, human review, or task
evaluation.
It can score random target-script text high, cannot separate shared-script
languages such as Pashto, Urdu, and Arabic, and may penalise valid code-switching.
Whisper tiny on Pashto illustrates a second failure mode: many baseline outputs
are romanized Pashto rather than arbitrary wrong-language text, while some
Arabic-script outputs still use the wrong shared-script variety.
For shared scripts, unique-glyph counts, LID when available, and human review are
needed after SFR.
Its positive use is to expose wrong-script failures hidden by WER tables; its
main risk is misuse as a sole quality gate.
Deployment use should combine SFR with lexical accuracy metrics and human review.

% §6 Conclusion — ~0.3 pages

\section{Conclusion}
\label{sec:conclusion}

Script Fidelity Rate (SFR) is a reference-free check for whether ASR output uses
the target writing system.
Across 100 FLEURS model--language pairs, 21 collapse: 20 Whisper pairs and one
Gemma~4 Urdu pair; MMS-1B and SeamlessM4T-v2 show no collapse.
For Gemma~4, script-aware prompting raises mean SFR from 71.2\,\% to 97.7\,\%;
for the six low-SFR languages, downstream NLLB translation recovers 5.9 chrF and
3.1 BLEU on average.

\printbibliography

% Appendix, not counted toward the 9-page NeurIPS content limit.

\appendix

\section{Full SFR matrix}
\label{app:matrix}

Table~\ref{tab:results} reports the numeric values visualised in
Figure~\ref{fig:sfr_heatmap}.
Cells with SFR below 10\,\% are \textbf{bold}; intermediate values
(10--90\,\%) are italicised.

\begin{table*}[htb]
\centering
\caption{Script Fidelity Rate (SFR, \%) for all 100 evaluated model--language
  pairs on FLEURS test sets.}
\label{tab:results}
\resizebox{\linewidth}{!}{%
\setlength{\tabcolsep}{3.5pt}
\renewcommand{\arraystretch}{1.05}
\begin{tabular}{lrrrrrrrrrr}
\toprule
& \multicolumn{4}{c}{Perso-Arabic}
& \multicolumn{4}{c}{Indic + Tamil (collapse targets)}
& Latin & Georgian \\
\cmidrule(lr){2-5}\cmidrule(lr){6-9}\cmidrule(lr){10-10}\cmidrule(lr){11-11}
Model       & Pashto & Urdu & Arabic & Persian & Hindi & Bengali & Malayalam & Tamil & Somali & Georgian \\
\midrule
Whisper tiny     & \textbf{7}   & 100 & 100 & 100 & \textbf{1}  & \textbf{0}  & \textbf{0}  & 100 & 99  & \textbf{0}   \\
Whisper base     & \textit{43}  & 100 & 100 & 100 & \textbf{5}  & \textbf{0}  & \textbf{0}  & 99  & \textbf{0}   & \textbf{0}   \\
Whisper small    & 97           & 100 & 99  & 100 & 100 & \textbf{0}  & \textbf{0}  & 99  & \textbf{0}   & \textit{16}  \\
Whisper medium   & \textit{34}  & 99  & 99  & 99  & 98  & \textbf{3}  & \textbf{2}  & 99  & \textbf{2}   & 94  \\
Whisper large-v2 & \textit{71}  & 100 & 99  & 99  & 96  & \textbf{1}  & \textbf{0}  & 99  & \textbf{0}   & \textit{41}  \\
Whisper large-v3 & 100          & 100 & 99  & 100 & 99  & 96          & \textbf{1}  & 99  & 100 & 100  \\
Whisper turbo    & 99           & 100 & 99  & 99  & 98  & 95          & \textit{13} & 99  & \textit{82}  & 97   \\
MMS-1B           & 99           & 99  & 100 & 100 & 99  & 99          & 99          & 99  & 99  & 99   \\
SeamlessM4T-v2   & 100          & 100 & 100 & 100 & 100 & 100         & 100         & 100 & 100 & 100  \\
\midrule
Gemma 4 E2B      & \textit{77} & \textbf{6} & 97 & 95 & 94 & \textit{89} & \textit{64} & \textit{70} & 100 & \textit{20} \\
\bottomrule
\end{tabular}}%
\end{table*}

\section{WER vs SFR scatter plot}
\label{app:scatter}

\begin{figure*}[htb]
  \centering
  \includegraphics[width=0.95\linewidth]{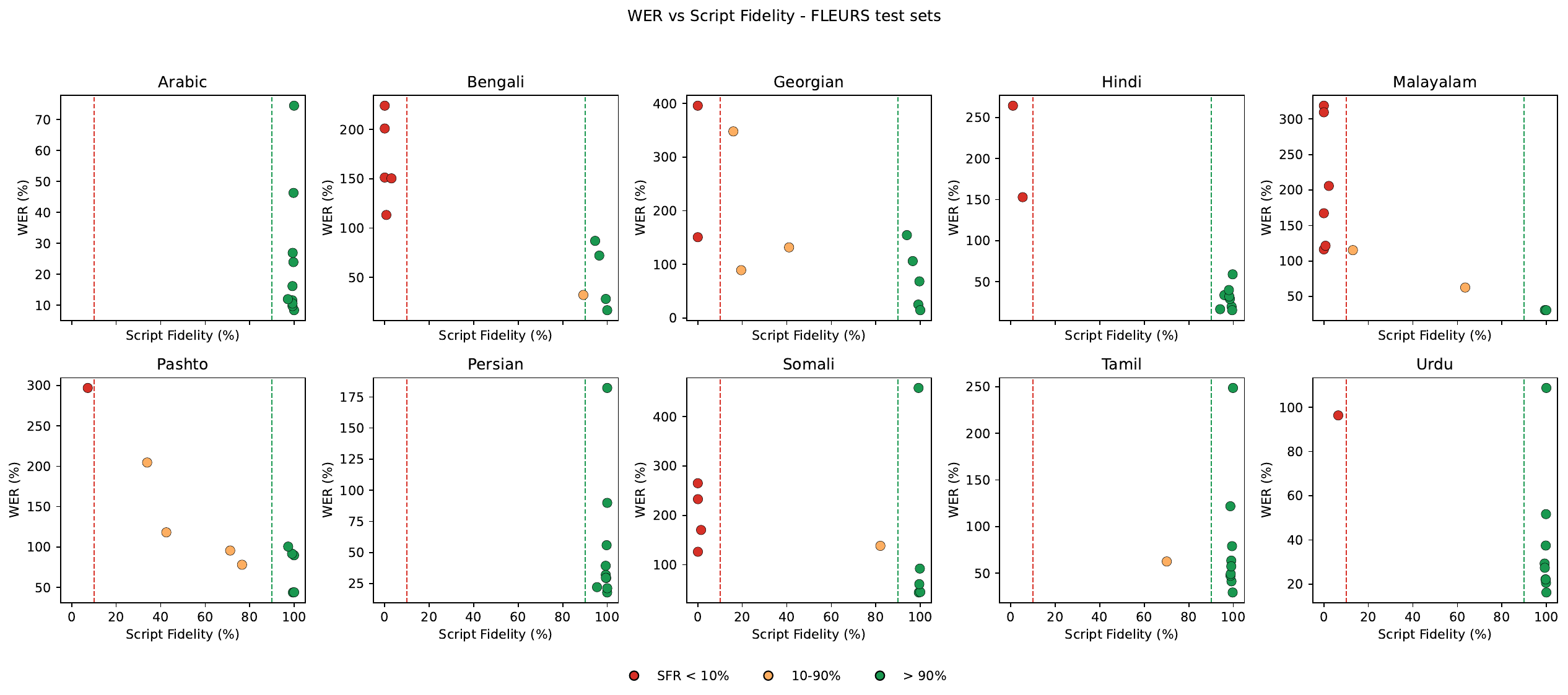}
  \caption{WER (\%) vs SFR (\%) for all 100 model--language pairs on FLEURS test
    sets, grouped by language.
    Red dashed line at SFR\,=\,10\,\% (collapse boundary); green dashed line at
    SFR\,=\,90\,\%.
    Points are coloured red (SFR $<$ 10\,\%), orange (10--90\,\%), and green
    ($>$ 90\,\%).}
  \label{fig:wer_sfr_scatter}
\end{figure*}

Figure~\ref{fig:wer_sfr_scatter} plots WER against SFR for all 100 model--language
pairs on FLEURS test sets.
Script collapse (low SFR, left region) is invisible to WER: the same WER value
can correspond to correct-script output or wrong-script output.
The Georgian panels show the widest SFR spread of any language, with Whisper
sizes ranging from 0\,\% (tiny, base) to 99.6\,\% (large-v3).

\section{Threshold sensitivity}
\label{app:threshold_sensitivity}

\begin{table}[htb]
\centering
\caption{Sensitivity of the collapse count to the SFR threshold.}
\label{tab:threshold_sensitivity}
\begin{tabular}{lrrrr}
\toprule
SFR threshold & 1\,\% & 5\,\% & 10\,\% & 20\,\% \\
\midrule
Collapsed pairs & 15 & 18 & 21 & 24 \\
\bottomrule
\end{tabular}
\end{table}

\section{Georgian SFR by model version}
\label{app:georgian}

\begin{figure}[htb]
  \centering
  \includegraphics[width=\linewidth]{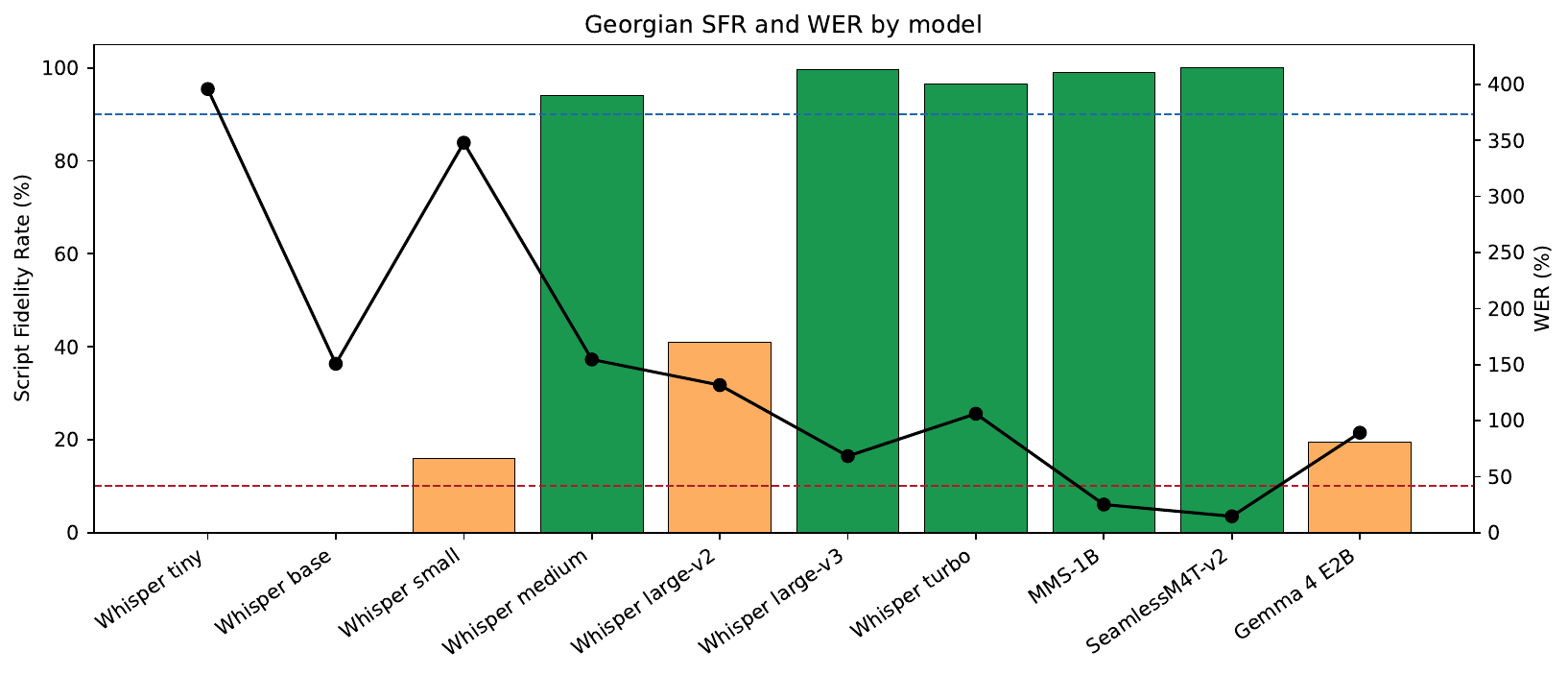}
  \caption{Georgian SFR (\%) and WER (\%) by model.
    SFR bars are coloured by collapse zone (red $<$ 10\,\%, orange 10--90\,\%,
    green $>$ 90\,\%).
    WER is plotted on the right axis.}
  \label{fig:georgian_detail}
\end{figure}

Figure~\ref{fig:georgian_detail} shows SFR and WER for Georgian, ordered by
Whisper model size.
The non-monotonic recovery between Whisper large-v2 (SFR\,=\,41\,\%) and
large-v3 (SFR\,=\,99.6\,\%) appears as a sharp step, while WER remains high for
both models.

\section{Taxonomy details}
\label{app:taxonomy_details}

\subsection{Dominant-script taxonomy table}
\label{app:taxonomy_table}

Table~\ref{tab:taxonomy} shows the breakdown for collapse-target languages
grouped by substitute-script pattern.
MMS-1B and SeamlessM4T-v2 produce target-script output at 99\,\% or above on
every language and are omitted from the table.

\begin{table*}[htb]
\centering\small
\caption{Dominant output script for collapse-target languages, grouped by
  pattern.
  SFR\,=\,Script Fidelity Rate (\%); Target\,=\,\% utterances with
  target-script-dominant output; Subst.\,=\,\% utterances with substitute
  script dominant.
  Rows omitted where SFR\,$\geq$\,90\,\%.
  ``Other'' covers mixed-script and unclassified blocks; rows may not sum to
  100.}
\label{tab:taxonomy}
\setlength{\tabcolsep}{5pt}
\begin{tabular}{llrrlr}
\toprule
Language & Model family & SFR & Target & Substitute script & Subst. \\
\midrule
\multicolumn{6}{l}{\textit{Pattern 1: Latin phonetic substitution}} \\
Pashto    & W-tiny/base      & 25 & 25 & Latin      & 69\,\% \\
Pashto    & W-small/med      & 66 & 64 & Latin      &  6\,\% \\
Pashto    & W-large-v2       & 71 & 71 & Latin      & 29\,\% \\
Hindi     & W-tiny/base      &  3 &  3 & Latin      & 82\,\% \\
Bengali   & W-tiny/base      &  0 &  0 & Latin      & 92\,\% \\
Malayalam & W-tiny/base      &  0 &  0 & Latin      & 70\,\% \\
\midrule
\multicolumn{6}{l}{\textit{Pattern 2: Arabic substitution}} \\
Somali    & W-tiny/base      & 50 & 48 & Arabic     & 50\,\% \\
Somali    & W-small/med      &  1 &  1 & Arabic     & 95\,\% \\
Somali    & W-large-v2       &  0 &  0 & Arabic     &100\,\% \\
\midrule
\multicolumn{6}{l}{\textit{Pattern 3: Devanagari substitution}} \\
Bengali   & W-small/med      &  2 &  1 & Devanagari & 58\,\% \\
Bengali   & W-large-v2       &  1 &  1 & Devanagari & 92\,\% \\
Malayalam & W-small/med      &  1 &  1 & Devanagari & 26\,\% \\
Malayalam & W-large-v2       &  0 &  0 & Devanagari & 23\,\% \\
Malayalam & W-large-v3       &  1 &  1 & Devanagari & 90\,\% \\
Malayalam & W-turbo          & 13 & 14 & Devanagari & 54\,\% \\
\midrule
\multicolumn{6}{l}{\textit{Pattern 4: Unique-script Latin collapse}} \\
Georgian  & W-tiny/base      &  0 &  0 & Latin      & 86\,\% \\
Georgian  & W-small/med      & 55 & 54 & Latin      & 13\,\% \\
Georgian  & W-large-v2       & 41 & 41 & Latin      & 57\,\% \\
\bottomrule
\end{tabular}
\end{table*}

\paragraph{Latin phonetic substitution.}
Whisper tiny and base often romanize non-Latin audio into phonetically plausible
Latin.
On Bengali, Whisper base routes 92\,\% of utterances to Latin-dominant output
(e.g.\ ``\textit{Jarmaneer on ek bekkara khabar}'' for a Bengali sentence).
The same pattern appears on Pashto and Malayalam.

\paragraph{Arabic substitution.}
Whisper small, medium, and large-v2 output Arabic-dominant text for Somali audio
despite Somali's Latin orthography since 1972.
This may reflect Arabic-script Somali text in the model's weakly supervised
training mix, but the public Whisper training data do not allow direct
verification.

\paragraph{Devanagari substitution.}
Several larger Whisper models output Devanagari for Bengali or Malayalam audio.
On Malayalam, Whisper large-v3 routes 90\,\% of utterances to
Devanagari-dominant output and Whisper turbo routes 54\,\%.
This pattern remains visible in the strongest Whisper releases evaluated here
and is absent in MMS-1B and SeamlessM4T-v2.

\paragraph{Unique-script Latin collapse.}
Whisper tiny and base output Latin for Georgian audio despite Georgian having no
script relationship to Latin.
Two sub-types appear in the hypotheses: phonetic approximations
(``\textit{Sogeerti cruisis pro sura zekermanis kalake berlinianachzen}'') and
unrelated English hallucinations (``\textit{Welcome to my channel}'').
The first sub-type resembles Latin phonetic substitution; the second suggests
fallback to a high-probability output distribution when the model lacks a
Georgian script prior.

\section{SFR vs language identification}
\label{app:lid}

Table~\ref{tab:sfr_lid} runs langdetect on saved Gemma~4 outputs.
The result supports a hybrid audit: SFR detects the wrong-script condition, while
LID can sometimes name the substitute language.
LID is not reliable enough to replace SFR.
It identifies Gemma~4 Urdu baseline output as Hindi, but labels high-SFR
Georgian script-hint output as Estonian and labels high-SFR Pashto as Persian
because Pashto is unsupported.

\begin{table}[htb]
\centering\small
\caption{SFR+LID audit on saved Gemma~4 predictions.
  LID uses langdetect; top label is the most frequent detected language.}
\label{tab:sfr_lid}
\setlength{\tabcolsep}{5pt}
\begin{tabular}{llrrlr}
\toprule
Language & Prompt & SFR & Low SFR & Top LID & Top LID \% \\
\midrule
Urdu     & baseline    &  6.5 & 88.6 & Hindi & 91.0 \\
Urdu     & script hint & 97.0 &  1.3 & Urdu  & 98.3 \\
Georgian & baseline    & 19.5 & 76.5 & Estonian & 27.3 \\
Georgian & script hint & 93.0 &  2.1 & Estonian & 83.0 \\
Pashto   & script hint & 98.8 &  0.2 & Persian & 99.2 \\
Somali   & baseline    & 99.6 &  0.0 & Somali & 99.7 \\
\bottomrule
\end{tabular}
\end{table}

\section{Deployment audit workflow}
\label{app:audit_workflow}

A minimal audit workflow requires no human annotation:
\begin{enumerate}
  \item Record ASR hypotheses and the intended target language.
  \item Compute utterance-level SFR using the target language's Unicode block
    specification.
  \item Alert when corpus-level mean SFR drops below a deployment threshold, for
    example $<0.8$.
  \item Inspect low-SFR examples with WER, CER, LID, or human review before making
    product decisions.
\end{enumerate}

\end{document}